\documentclass[12pt,a4paper]{article}
\pdfoutput=1

\usepackage{amsmath,amssymb,amsfonts,mathrsfs,mathtools,graphicx,bbm}
\usepackage{fixltx2e}
\usepackage[font=small]{caption}
\usepackage{lmodern}
\usepackage{booktabs}
\usepackage{dsfont}
\usepackage{slashed}
\usepackage{braket}
\usepackage{hyperref}
\usepackage{xspace}

\usepackage{epsfig}
\usepackage{relsize}
\input epsf
\usepackage{enumitem}
\usepackage{color}

\newcommand{\be}{\begin{equation}}
\newcommand{\ee}{\end{equation}}
\newcommand{\bea}{\begin{eqnarray}}
\newcommand{\eea}{\end{eqnarray}}
\newcommand{\bee}{\begin{enumerate}}
\newcommand{\eee}{\end{enumerate}}
\newcommand{\bei}{\begin{itemize}}
\newcommand{\eei}{\end{itemize}}
\newcommand{\bal}{\begin{equation}\begin{aligned}}
\newcommand{\eal}{\end{aligned}\end{equation}}
\newcommand{\bem}{\left (\begin{matrix}}
\newcommand{\eem}{\end{matrix} \right )}

\newcommand{\nn}{\nonumber}
\newcommand{\la}{\label}

\newcommand{\alg}[1]{\mathfrak{#1}}
\newcommand{\su}{\alg{su}}

\def\sl2{\alg{sl}(2)}

\newcommand{\ads}{${\rm  AdS}_5\times {\rm S}^5\ $}

\numberwithin{equation}{section}
\makeatletter
 \let\old@startsection=\@startsection
 \let\oldl@section=\l@section
 \renewcommand{\@startsection}[6]{\old@startsection{#1}{#2}{#3}{#4}{#5}{#6\mathversion{bold}}}
 \renewcommand{\l@section}[2]{\oldl@section{\mathversion{bold}#1}{#2}}
\makeatother

\DeclareMathOperator{\diag}{diag}

\newcommand{\AdS}{\text{AdS}}

\definecolor{grey}{rgb}{0.4,0.4,0.5}
\definecolor{darkgreen}{rgb}{0,0.5,0}
\definecolor{darkred}{rgb}{0.6,0.0,0}
\definecolor{lightbrown}{rgb}{1,0.9,0.8}
\definecolor{brown}{rgb}{0.6,0.3,0.3}
\definecolor{darkblue}{rgb}{0,0,0.5}
\definecolor{darkmagenta}{rgb}{0.5,0,0.5}

\definecolor{TCDblue}{rgb}{0.1,0.3,0.6}

\def\pa {\partial}
\def\ov{\over}

\def\a {\alpha}
\def\b {\beta}
\def\g {\gamma}
\def\G {\Gamma}
\def\vG {\varGamma}
\def\de {\delta}
\def\D {\Delta}
\def\De {\Delta}

\def\e{\epsilon}

\def\k{\kappa}

\def\m{\mu}
\def\r {\rho}

\def\s {\sigma}

\def\w {\omega}

\def\p{\phi}

\def\vp{\varphi}

\def\t{\theta}

\def\vt{\vartheta}

\def\u\upsilon
\def\U\Upsilon
\def\vU\varUpsilon
\def\vPi{\varPi}

\def\cE{{\cal E}}

\def\cH{{\cal H}}

\def\cL{{\cal L}}
\def\cM{{\cal M}}

\def\cO{{\cal O}}
\def\cP{{\cal P}}

\def\cR{{\cal R}}

\def\bZ{{\mathbb Z}}

\def\rA{{\rm A}}

\def\rF{{\rm F}}
\def\rG{{\rm G}}

\def\rK{{\rm K}}

\def\rV{{\rm V}}

\def\dx{{\dot x}}

\def\th{{\tilde h}}

\newcommand\bp{\hbox{\larger $\pi$}}

\def\x'{\mathaccent 19 x}
\def\y'{\mathaccent 19 y}
\def\n'{\mathaccent 19 n}
\def\u'{\mathaccent 19 u}

\def\et'{\mathaccent 19 \eta}
\def\th'{\mathaccent 19 \theta}
\def\lam'{\mathaccent 19 \lambda}
\def\varet'{\mathaccent 19 \vartheta}
\def\rh'{\mathaccent 19 \rho}
\def\ph'{\mathaccent 19 \phi}
\def\xb'{\mathaccent 19 {\bar{x}}}

\def\d1{{\dot{1}}}

\def\sun2{SU(N)$\times$SU(N)}

\def\TTb{${T\bar T}\ $}

\begin{document}

\null\vskip-40pt
 \vskip-5pt \hfill
\vskip-5pt \hfill {\tt\footnotesize
TCD-MATH-19-06}

\renewcommand{\thefootnote}{\fnsymbol{footnote}}

\vskip 1cm \vskip0.2truecm
\begin{center}
\begin{center}
\vskip 0.8truecm {\Large\bf \TTb deformation and the light-cone gauge\footnote[1]{Invited contribution to the special issue of the ``Proceedings of the Steklov Institute of Mathematics'' dedicated to the 80th anniversary of Andrei Slavnov.}
}
\end{center}

\vskip 0.9truecm
Sergey Frolov\footnote[2]{ Correspondent fellow at
Steklov Mathematical Institute, Moscow.}\footnote[3]{email: 
frolovs@maths.tcd.ie}
 \\
\vskip 0.5cm

{\it School of Mathematics and Hamilton Mathematics Institute, \\
Trinity College, Dublin 2,
Ireland}

\end{center}
\vskip 1cm \noindent\centerline{\bf Abstract} \vskip 0.2cm 

 The homogeneous inviscid Burgers equation which determines the spectrum of a \TTb deformed model has a natural interpretation as the condition of the gauge invariance of the target space-time energy and momentum of a (non-critical) string theory quantised in a generalised uniform light-cone gauge which depends on the deformation parameter. As a simple application of the light-cone gauge interpretation we derive the \TTb deformed Lagrangian for a system of any number of scalars, fermions and chiral bosons with an arbitrary potential. We find  that the \TTb deformation is driven by the canonical Noether stress-energy tensor but not the covariant  one.

\flushbottom

\newpage

\tableofcontents

\renewcommand{\thefootnote}{\arabic{footnote}}
\setcounter{footnote}{0}

\section{Introduction and summary}

This paper is dedicated to Andrei Alekseevich Slavnov, the best supervisor one can hope for, on the occasion of his 80th birthday. He taught me the methods used in this paper.

\medskip

The irrelevant deformation of a 2d field theory by the \TTb operator introduced in \cite{Z04} has attracted a lot of attention after the \TTb deformation of integrable field theories was analysed in \cite{SZ16, Tateo16}, for recent lecture notes see \cite{Jiang19}. The most interesting feature of a  \TTb deformed model is that its spectrum is completely fixed by the spectrum of the undeformed model \cite{Z04}. In the case where the momentum of a state is equal to 0, the energy of the state, as a function of the deformation parameter $\a$ and the circumference $R$ of the cylinder the theory is defined on, satisfies the homogeneous inviscid Burgers equation 
\bal\la{Bureqhomintro}\nn
\pa_\a \cE_\a(R)+\cE_\a(R)\pa_R\cE_\a(R)=0\,,
\eal
whose integrated form is
\bal\la{Bureqintro}
\cE_\a(R)=\cE_0(R-\a\, \cE_\a(R))\,.
\eal
Introducing the circumference $R_0$ of the undeformed theory  
\bal\nn
R_0=R-\a\, \cE_\a(R)\quad \Longleftrightarrow\quad R=R_0+\a\, \cE_0(R_0)\,,
\eal
the eq.\eqref{Bureqintro} takes the form
\bal\la{Bureqintro2}
\cE_\a(R_0+\a\, \cE_0(R_0))=\cE_0(R_0)\,.
\eal
Thus, the homogeneous inviscid Burgers equation is just a statement that the energy of a \TTb deformed theory on a circle of circumference $R_0+\a\, \cE_0(R_0)$ is independent of the deformation parameter. 
If we now denote
\bal\la{Bureqintro3}\nn
\cE_\a(R_0+\a\, \cE_0(R_0))=H=\int_0^{P_-}d\s\cH_{\rm ws} \,,\quad  R_0+\a\, \cE_0(R_0)=P_-=J+a H\,,
\eal
then we see that eq.\eqref{Bureqintro2} is the same as eq.(2.17) in the review \cite{AFrev}
where the propagation of strings on a background with time and space isometries was analysed in a so-called uniform light-cone gauge introduced in \cite{AFZ06a}. This is a one-parameter generalisation of the standard light-cone gauge which corresponds to $a=1/2$. Due to the two isometry directions,  
the target
space-time energy $E$ and the total momentum $J$ are conserved.  They are  independent of a gauge choice if the total world-sheet momentum vanishes, and since $H=E-J$ is the gauge-fixed light-cone world-sheet Hamiltonian  one immediately gets \eqref{Bureqintro2}.

\medskip

Thus, we conclude that the homogeneous inviscid Burgers equation which determines the spectrum of a \TTb deformed model at vanishing  world-sheet momentum can be interpreted as the condition of  gauge invariance of the target space-time energy and momentum of a (non-critical) string theory quantised in a  uniform light-cone gauge. It is also clear that the deformation parameter $\a$ should be related to the gauge parameter $a$ as $a=1/2+\a$ because for $a=1/2$ the light-cone strings in flat space are described by a free theory which is naturally taken as the undeformed model. The light-cone strings in \ads space are not described by a free theory for any choice of the gauge parameter, and the most natural choice is $a=0$ which explains the parametrisation chosen in \cite{AFZ06a}. 

\medskip

Let us stress that the world-sheet Hamiltonian density $\cH_{\rm ws}$ does depend on a gauge parameter in a very nontrivial way, and if one fixes the world-sheet size $R$ then the spectrum of the 
world-sheet Hamiltonian $H$ will depend on the gauge parameter too. That is why a gauge parameter  can be treated as a deformation one. 

\medskip

If the world-sheet momentum does not vanish the target space-time energy $E$ and momentum $J$ are not gauge-invariant anymore. Still, in the case of Lorentz invariant models 
the relation to a light-cone gauge-fixed string sigma model can be used to derive the inhomogeneous inviscid Burgers equation governing the spectrum for any world-sheet momentum as we demonstrate in section \ref{Burgers}.

\medskip

The light-cone gauge interpretation gets more support if we note that the CDD factor $e^{-i\a m^2\sinh\t}$ which relates the deformed and undeformed models exactly coincides with the $a$-dependent factor in eq.(8.9) of \cite{AFZ06b}, see also eq.(3.94) of \cite{AFrev}.\footnote{The derivation of the factor was not given in \cite{AFZ06b} due to its simplicity but it will be reviewed in subsection \ref{Burgers} for reader's convenience.} 
Indeed, introducing the rapidities $\t_1$ and $\t_2$ of the colliding particles, the \TTb CDD factor  can be rewritten in the form
\bal\la{CDDf}
e^{-i\a m^2\sinh(\t_1-\t_2)}=e^{-i\a(p_1\w_2-p_2\w_1)}\,,\quad p_k=m \sinh\t_k\,,\quad \w_k=m \cosh\t_k\,,
\eal
used in \cite{AFZ06b}.  The \TTb CDD factor also appeared in the study of effective bosonic string theory in flat space in \cite{Dubovsky12}, and its relation to the \TTb deformation was noticed in
\cite{Tateo13}.  Recently, it was pointed out that it also describes the world-sheet scattering of light-cone strings on AdS$_3$ backgrounds without RR fields
\cite{Sfondrini18, Sfondrini18a,Sfondrini18c}.  It is worthwhile mentioning that in the form \eqref{CDDf} the \TTb CDD factor is also valid for massless particles and only affects the left-right scattering \cite{Sfondrini18, Sfondrini18a,Sfondrini18c}, and for  a nonrelativistic (integrable) model with any dispersion relation.

\medskip

It is often said that the \TTb deformation of free bosons is flat space string theory in static gauge with a deformation parameter dependent $B$ field. This is technically correct but it hides the actual origin of the $B$ field, and it provides no insight on how to find a \TTb deformed action. It is known \cite{KT04,Z09} that a uniform light-cone gauge-fixed action can be obtained by first T-dualising in the $x^-$ direction,\footnote{See eq.\eqref{lcc} for the definition of $x^\pm$.} then integrating out the world-sheet metric, and finally fixing the  static gauge, $x^+=\tau$, $\tilde x^-=\s$ in the resulting Nambu-Goto action. The $B$ field appears as a result of the T-duality but there is no $B$ field in the string sigma model we start with.  We do not see much (if any) technical advantages in performing the T-duality, and in this paper we only use the standard phase space approach to a light-cone gauge fixing which provides a crystal clear relation to the Burgers equation.

\medskip

The relation of  string theory in the uniform light-cone gauge to the \TTb deformation of conformal field theories is not new. It was mentioned in \cite{Sfondrini18},\footnote{The interpretation was also known to the author \cite{SFjournal}.} and was successfully used to analyse \TTb deformation of free supersymmetric models in \cite{Sfondrini18b,Sfondrini19a}. Apparently, the full power of the light-cone gauge approach has not been fully appreciated by the \TTb deformed community \cite{Sfondrinitalk}. 

\medskip

In this paper we generalise this approach to any model and, as a simple application of the light-cone gauge interpretation, we derive the \TTb deformed Lagrangian for a (not necessarily Lorentz invariant) system of any number of scalars, fermions and chiral bosons with an arbitrary potential, see eq.\eqref{LfermTTb}. 
We find  that the \TTb deformation is always driven by the canonical Noether stress-energy tensor but not the covariant  tensor (obtained by the metric variation), see \eqref{Tcanf}. A similar observation was made in \cite{Sfondrini18b,Sfondrini19a}. It is natural because for a non-Lorentz invariant model the covariant stress-energy tensor may not be defined.

\medskip

The  \TTb deformed Lagrangian \eqref{LfermTTb} of a model with bosons is always of the square root Nambu-Goto form. However, it simplifies drastically for a purely fermionic model
of $n_f^-$ right-moving real fermions $\t_-^i$, and $n_f^+$ left-moving real fermions $\t_+^r$, and is given by
 \bal\la{LfermTTbintro}
\cL_{\rm AAF}&={\, i\rK^+_++i\rK^-_-+\a(\rK^+_-\rK^-_+-\rK^+_+\rK^-_-)\, - \rV \ov 1+\a\rV }\,.
\eal 
Here
\bal
\rK^+_\g\equiv \t_-^i\rK_{ij}^+\pa_\g\t_-^j\,,\quad \rK^-_\g\equiv\t_+^r\rK_{rs}^-\pa_\g\t_+^s\,,\quad \pa_\pm=\pa_\tau\pm\pa_\s\,,
\eal
$\rK_{ab}^\pm$ are fermion kinetic matrices whose $\t$-independent pieces are symmetric and can be diagonalised, and 
 $\rV$ is an arbitrary potential. If we want a Lorentz invariant model, then $\rK_{ab}^\pm$ and $\rV$ may depend on the products $\t_-^i\t_+^r$ only.

\medskip

The Lagrangian \eqref{LfermTTbintro} is in fact a generalisation of the Alday-Arutyunov-Frolov (AAF) model \cite{AAF} to any number of  fermions and any potential. The AAF model  is an integrable model of a massive Dirac fermion with $\a=-1/2$. It describes the $\su(1|1)$ sector of the \ads superstring in the $a=0$ light-cone gauge. In the modern terminology the AAF model is the \TTb deformation of a free massive Dirac fermion.  In fact it was obtained in \cite{AAF} in exactly the same way as the one described in section \ref{TTbfermions} of this paper. Some properties of the model were investigated in \cite{KZ06}-\cite{Melikyan16}, and we will discuss the implications of these studies in Conclusions. It was later realised in \cite{AFlc} that quantising the $\su(1|1)$ sector in the standard $a=1/2$ light-cone gauge leads to a free massive Dirac fermion, and the independence of the target space-time energy $E$ and momentum $J$ of a gauge choice  was used to find the spectrum of $E$ in the semi-classical approximation which was the quantity of interest for the AdS/CFT correspondence \cite{M}. 

\medskip

Let us also mention that if we are not interested in Lorentz invariance then $\rK_{ab}^\pm$ and $\rV$ can have any dependence on the fermions. In particular, if we only consider the right-moving fermions
and choose for simplicity $\rK_{ij}^+=\delta_{ij}$ then \eqref{LfermTTbintro} takes the following form
\bal\la{LfermTTbintro2}
\cL_{\rm SYK}&={\, i\t_-^k\pa_+\t_-^k\, - \rV \ov 1+\a\rV }\,.
\eal
Forgetting about  $\s^-$ coordinate, one can interpret this Lagrangian as a \TTb deformation of the SYK model \cite{SYK1,SYK2}. It would be interesting to see how the properties of the SYK model are modified by the \TTb deformation.

\medskip

The plan of the paper is as follows. In section 2.1 we review the construction of the uniform light-cone gauge-fixed action for bosonic strings propagating in a target manifold possessing
 time and  space abelian isometries. In section 2.2 we explain how the inviscid Burgers equation and the \TTb CDD factor appear in the light-cone gauge approach. In section 2.3 the \TTb deformation of a Lorentz invariant sigma-model  of  bosonic fields with an arbitrary potential is considered. In section 3.1 we generalise the consideration in section 2.1 to a Green-Schwarz type sigma model with any number of scalars, fermions and chiral bosons. In section 3.2  the \TTb deformed Lagrangian for a system of any number of scalars, fermions and chiral bosons with an arbitrary potential is derived. Finally, in Conclusions we discuss open questions and generalisations of the light-cone gauge approach.

\section{Light-cone gauge and inviscid Burgers equation}\la{blc}
\subsection{Uniform light-cone gauge}\la{unlc}

In this subsection we follow  closely the review \cite{AFrev}. To explain the ideas in this subsection we only consider bosonic strings propagating in a $n+2-$dimensional target Minkowski manifold $\cM$ possessing
(at least) two abelian isometries, one of which is in the time direction.  
We denote coordinates of $\cM$ by $X^M$, $M=0,1,\ldots,n+1$, the  time and space isometry
coordinates  by $t\equiv X^0 $ and  $\phi\equiv X^{n+1}$, respectively, and
 the ``transversal'' coordinates by $x^\mu\equiv X^\mu$, $\mu=1,\ldots,n$.
The two abelian isometries are realised by shifts of $t$ and $\phi$.
If the variable $\phi$ is an angle then the
range of $\phi$ is  from $0$ to $2\pi R_{\phi}$. Obviously, the metric $G_{MN}$ of $\cM$ does not depend on  $t$ and $\phi$. We assume for simplicity that the components $G_{t\mu}$ and $G_{\p\mu}$ vanish.
Thus, the metric  of $\cM$ is of the form
\bea
\la{metrgen} ds^2=G_{MN}dX^MdX^N = G_{tt}\, dt^2\, +\, 2G_{t\p}\, dtd\p\, +\,G_{\p\p}\, d\p^2\, +\,
G_{\mu\nu}\,dx^\mu dx^\nu \, ,\eea 
where
$G_{MN}$ is the target-space metric independent of $t$ and $\p$.

\medskip

We  assume that the $B$-field vanishes but it can be easily included if necessary, see e.g. \cite{AT14}.\footnote{If the $B$-field has nonvanishing components only in the transverse directions $x^\mu$ then the analysis below is not really modified. }
Then, the string action is given by 
\bea \la{S1} 
S = -{1\over 2}\int_{-r}^{ r}\, {\rm
d}\s{\rm d}\tau\, \g^{\a\b}\partial_\a X^M\partial_\b X^N\,
G_{MN}\,, 
\eea 
where $\gamma^{\a\b}= h^{\a\b} \sqrt {-h}$ is the Weyl-invariant
combination of the world-sheet
metric $h_{\a\beta}$ with $\det\gamma=-1$.\footnote{In the conformal gauge
$\g^{\a\b} = {\mbox{diag}}(-1,1)$.} The range of the world-sheet space coordinate $\s$ is $-r\le\s\le r$ where $r$ will be fixed by a generalised uniform light-cone gauge.

\medskip

To impose a uniform light-cone gauge  we transform the
string action (\ref{S1}) to the first-order form 
\bea \la{S2} 
S=\int_{- r}^{ r}\, {\rm d}\s{\rm d}\tau\, \left( p_M \dot{X}^M +
{\g^{01}\ov\g^{00}} C_1+ {1\ov 2\, \g^{00}}C_2\right)\,.  
\eea
Here $p_M$ are momenta canonically-conjugate to the coordinates $X^M$
 \bea\nonumber 
 p_M ={\de S\ov \de \dot{X}^M} = -\g^{0\b}\partial_\b X^N\,
G_{MN}\,,\quad \dot{X}^M\equiv \pa_0 X^M \,, 
\eea and
$C_1$ and $C_2$ are the
two Virasoro constraints 
\bea \nonumber 
C_1=p_MX'^M\,,\quad
C_2=G^{MN} p_M p_N + X'^M X'^N G_{MN}\,,\qquad X'^M\equiv
\pa_1 X^M\,, 
\eea 
which are solved after imposing a light-cone gauge.

\medskip

The  string action invariance under the shifts of $t$ and $\p$
implies the conservation of the target
space-time energy $E$, and of the total (angular) momentum $J$ of the string in the
$\p$-direction
\bea\la{charges} 
E = - \int_{-
r}^{ r}\, {\rm d}\s\, p_t\ \, , \qquad J= \int_{- r}^{ r}\, {\rm
d}\s\, p_\p\ . 
\eea 
It is clear that the charges $E$ and $J$ are gauge-independent.

\medskip

To impose a uniform gauge we introduce the ``light-cone''
coordinates and momenta: 
\bal \la{lcc} 
&x^- =\p \,-\,t\ , \,\,
x^+ ={1\ov2}(\p\, +\,t)+\a\, x^-\ ,\,\, p_+ = p_\p\,+\,p_t\ ,\,\, p_- =
{1\ov2}(p_\p\, -\,p_t)-\a\, p_+\, .
 \eal 
 Here $\a$ is an arbitrary parameter\footnote{As was mentioned in the Introduction, it is related to the parameter $a$ used in \cite{AFZ06a,AFrev} as
 $a={1\ov2}+\a$. We have also slightly changed the notations in comparison to \cite{AFrev}: $x^\pm=x_\pm^{\rm there}$, $p_\pm=p_\mp^{\rm there}$, to make them closer to the commonly used. } of the most general light-cone coordinates such
that the light-cone momentum $p_+$ is equal to $p_+=p_\p\,+\,p_t$. As a result, in
 the corresponding uniform light-cone gauge
 the world-sheet
Hamiltonian is equal to $H_{\rm ws}=E-J=-P_+$. As we will show, $\a$ is a \TTb deformation parameter used in \cite{SZ16}. 

\medskip

The total light-cone momenta are found by using (\ref{charges})
\bea\la{charges2}\nonumber
P_+ \,=\,\int_{- r}^{ r}\, {\rm d}\s\, p_+\,= \, J\,-\,E\ \, ,
\qquad P_- \,=\, \int_{- r}^{ r}\, {\rm d}\s\, p_-\,=\, {1\ov2}(J\, +\, E)-\a\, P_+\, . 
\eea 
Assuming that the target space-time metric
is of the form (\ref{metrgen}), and by using the light-cone coordinates we write the
action (\ref{S2}) in the form 
\bea \la{S3} 
S = \int_{- r}^{r}\, {\rm d}\s{\rm d}\tau\, \left( p_+\dot{x}^++ p_- \dot{x}^- +
p_\mu \dot{x}^\mu + {\g^{01}\ov\g^{00}} C_1+ {1\ov 2\,
\g^{00}}C_2\right)\, , 
\eea 
where 
\bea 
\la{C1b} C_1\,=\,p_+x'^+
\,+\, p_-x'^- \,+\, p_\mu x'^\mu\,  ,
\eea 
and
\bal \la{c2}
C_2 &= G^{++} p_+^2\,  +\, 2G^{-+}p_-p_+ \, +\,
G^{--}p_-^2\\ &+
G_{++} (x'^+)^2\,  +\,
2G_{-+} x'^-x'^+
+ G_{--}(x'^-)^2 \, + \,
2\cH_x\,.
\eal 
Here $\cH_x$ is the part of the constraint which
depends only on the transversal fields $x^\mu$ and $p_\mu$
\bea\la{Hx} 
\cH_x = {1\ov2}\big(G^{\mu\nu}p_\mu p_\nu +  x'^\mu x'^\nu\,
G_{\mu\nu}\big)\,, 
\eea 
and the light-cone components of the target space metric are given by
\bal\la{Gpp}
G_{++}=G_{\p\p}+2 G_{t\p}+G_{tt}
\,,\quad
G^{--}={G_{++}\ov \det G_{\rm lc}}\,,\quad \det G_{\rm lc}\equiv G_{tt} G_{t\p} - G_{\p\p}^2
\,,
\eal 
\bal\la{Gmm}
G_{--}=\left(\frac{1}{2}-\alpha\right)^2 G_{\p\p}+\left(2 \alpha ^2-\frac{1}{2}\right) G_{t\p}+\left(\frac{1}{2}+\alpha \right)^2 G_{tt}\,,\quad
G^{++}={G_{--}\ov \det G_{\rm lc}}\,,
\eal 
\bal\la{Gmp}
G_{-+}=\left(\frac{1}{2}-\alpha\right) G_{\p\p}-2 \alpha  G_{t\p}-\left(\alpha +\frac{1}{2}\right) G_{tt}\,,\quad
G^{-+}=-{G_{-+}\ov \det G_{\rm lc}}\,.
\eal 

\medskip

A uniform light-cone gauge is defined by the two
conditions 
\bal \la{ulc} x^+ \,=\, \tau \,+\, a\,{\pi\ov r} mR_\p\,\s\
,\quad p_- \,=\, 1\,,\qquad a = {1\ov2}+\a\,,
\eal
 where $m$ is an integer winding number which
represents the number of times the string winds around the circle
parametrised by $\phi$. It appears if $\p$ is an angle
variable with the range $0\le\p\le 2\pi R_\p$ and, therefore, it obeys the constraint
 \bea \label{period}
\phi(r)-\phi(-r)=2\pi  m R_\p\, , ~~~~~m\in {\mathbb Z}\, . 
\eea 
Integrating the gauge condition $p_-=1$ over $\s$, we 
relate  the constant $r$ to
the total light-cone momentum
\bal\nonumber r = {1\ov 2}P_-\,.
\eal
Thus, the world-sheet of the
light-cone string  model is a cylinder of
circumference $P_-$.

\medskip

The gauge-fixed action is found by solving the Virasoro
constraints. First, we use $C_1$ to find $x'^-$ 
\bal\la{C1c}
 C_1\,=\,x'^- \,+\,  a {2\pi\ov P_-} mR_\p\, p_+ \,+\,
p_\mu x'^\mu\,=\,0\, \quad \Rightarrow \quad x'^-\,=\, - a {2\pi\ov P_-} m
R_\p\, p_+ \,-\, p_\mu x'^\mu\,.
\eal
Then, the solution is substituted 
into $C_2$,  and the resulting equation $C_2=0$ is solved for $p_+$.
Finally, having found these solutions we bring the string action
(\ref{S3}) to the gauge-fixed form
\bea \la{S4} 
S_\a=\int_{- r}^{ r}\, {\rm d}\s{\rm d}\tau\, \left( p_\mu \dot{x}^\mu
\,-\, \cH_{\rm ws} \right)\, ,
 \eea
  where 
  \bal \la{denH} 
  \cH_{\rm ws} \,=\, -p_+(p_\mu, x^\mu ,x'^\mu )\,, 
  \eal 
  is the
density of the world-sheet Hamiltonian which depends only on the
transversal
 fields $p_\mu, x^\mu$ which are periodic, $x^\mu(r) = x^\mu(-r)$, because we assumed that the strings are closed. Thus, the gauge-fixed string
action describes a two-dimensional model on a cylinder of
circumference $2 r =P_-$. Clearly, for generic values of $\a$ the gauge-fixed world-sheet Hamiltonian is of a square root Nambu-Goto type.
The two-dimensional model is not in general Lorentz invariant on the world-sheet. However, it is  invariant under the shifts of
the world-sheet coordinate $\s$, and therefore,  the total
world-sheet momentum of the string  is conserved
\bea\la{pws} 
P_{{\rm ws}} = -\int_{- r}^{
r}\, {\rm d}\s\, p_\mu x'^\mu\,. 
\eea 
States of the resulting two-dimensional model may have any world-sheet momentum. However, only the physical states which satisfy the level-matching condition 
\bal\la{LM} \De x^-=
\int_{- r}^{ r}\, {\rm d}\s\, x'^- = a{2\pi\ov P_-} mR_\p\, H_{\rm ws} - \int_{- r}^{ r}\, {\rm
d}\s\, p_\mu x'^\mu = a{2\pi\ov P_-} mR_\p\, H_{\rm ws} +P_{\rm ws}= 2\pi mR_\p\,, 
\eal 
have the target space-time energy $E$  and momentum $J$, and therefore the world-sheet energy $E_{\rm ws}=E-J$, independent of the gauge parameter $\a$. Solving eq.\eqref{LM} for $P_{\rm ws}$, one finds
\bal\la{pws2} 
P_{\rm ws}={2\pi m R_\p(P_--aH_{\rm ws})\ov P_-}={2\pi m R_\p J\ov P_-}={2\pi m k\ov P_-}\,,
\eal
where we have taken into account that in quantum theory the charge $J$ should be quantised: 
$J =k/R_\p$, $k\in\bZ$. 

\subsection{Inviscid Burgers equation}\la{Burgers}

Now we are ready to derive the inviscid Burgers equation. 
We consider a physical state with momentum $P_{\rm ws}$ given by \eqref{pws2}, and energy $E_{\rm ws}(R,\a)$ of a light-cone gauge-fixed model on a cylinder of circumference $P_-$.
To simplify the notations we denote
$R\equiv P_-$, and  $\cE_\a(R)\equiv E_{\rm ws}(R,\a)$. Then, we have
\bal\la{Bureq0}
&\cE_\a(R)=E-J=\cE_0(R_0)\,,\quad R_0\equiv {1\ov2}(J\, +\, E)\,,\\ 
R=R_0+&\a\,\cE_\a(R)=R_0+\a\,\cE_0(R_0)\quad \Longleftrightarrow\quad R_0=R-\a\, \cE_\a(R)   \,.
\eal
Thus,
\bal\la{Bureq1}
\cE_\a(R)=\cE_0(R_0)=\cE_0\big(R-\a\, \cE_\a(R) \big)\,,
\eal
which is the integrated form of the homogeneous inviscid Burgers equation
\bal\la{Bureqhom}
\pa_\a \cE_\a(R)+{1\ov 2}\pa_R \cE^2_\a(R)=0\,,
\eal
and $\a$ is indeed equal to the \TTb deformation parameter used in \cite{SZ16}.

\medskip

It is worth noting that if  $m=0$  then $\cH_{\rm ws}$ has no
dependence on $P_-$, and the dependence of the gauge-fixed 
 world-sheet Hamiltonian $H_{\rm ws} \,=\, \int_{- r}^{ r}\, {\rm
d}\s\, \cH_{\rm ws}$ on $P_-$ comes only through the integration bounds $\pm
r$. In this situation we can consider the decompactification  limit where
$P_-=R\to\infty$, and get a two-dimensional
 model defined on a plane. Let us assume that the asymptotic states and S-matrix are
well-defined. To  find a relation between the deformed and undeformed S-matrices  let us also assume that the models are integrable. Then, at large $R$ the spectrum of these models is determined by Bethe equations\footnote{We assume for simplicity that the scattering matrices are diagonal. In general, the spectrum would be described by a nested Bethe ansatz. Since Bethe equations for auxiliary roots do not depend on $R$, they do not change the conclusion.}
\bal\la{BE1}
e^{ip_i R}\prod_{k\neq i}^N S_{ik}(p_i,p_k)&=1\,,\quad \sum_{k=1}^N p_k=0\,,
\eal
\bal
\cE_\a(R) = \sum_{i=1}^N\w_i\,,\quad R \gg1\,,
\eal
where $\w_i$ is the dispersion relation of the $i$-th particle.  
Since  
$R=R_0+\a\, \cE_\a(R)$ we can rewrite \eqref{BE1} in the form
\bal\la{BE2}
e^{ip_i R_0}\prod_{k\neq i}^N e^{i\a(p_i\w_k-p_k\w_i)}S_{ik}(p_i,p_k)&=1\,,\quad \sum_{k=1}^N p_k=0\,.
\eal
These are Bethe equations of the undeformed model with the S-matrices related as\footnote{For a nonrelativistic model the dispersion relations of deformed and undeformed models do not have to coincide.}
\bal\la{SSTTb}
S_{ik}^{(0)}(p_i,p_k)=e^{i\a(p_i\w_k-p_k\w_i)}S_{ik}(p_i,p_k)\,,\quad S_{ik}(p_i,p_k)=e^{-i\a(p_i\w_k^{(0)}-p_k\w_i^{(0)})}S_{ik}^{(0)}(p_i,p_k)\,.
\eal
The CDD factor of the form $e^{-i\a(p_i\w_k-p_k\w_i)}$ appeared explicitly\footnote{Its existence was also mentioned in \cite{FPZ06,AF06}, and it was computed perturbatively to leading order in small momentum expansion in \cite{KMRZ}.} in eq.(8.9) of \cite{AFZ06b}
in the study of the \ads world-sheet S-matrix. For a relativistic theory one has
\bal
\w_i=m_i\cosh\theta_i\,,\quad p_i=m_i\sinh\theta_i\,,\quad p_i\w_k-p_k\w_i=m_im_k\sinh(\theta_i-\theta_k)\,,
\eal
and one reproduces the phase found in \cite{SZ16}
\bal
S_{ik}(\theta_{ik})=e^{-i\a m_im_k\sinh\theta_{ik}}S_{ik}^{(0)}(\theta_{ik})\,,\quad \theta_{ik}=\theta_i-\theta_k\,.
\eal
The relations \eqref{SSTTb} between the deformed and undeformed S-matrices are valid for non-integrable models if the energies of scattering particles are low enough. It has been argued in \cite{Dubovsky17} that these relations are exact and valid for any model and arbitrary energies.
It would be interesting to see if one could come to the same conclusion in the current approach.

\medskip

 Eq.\eqref{Bureq1} is valid only for physical states with the world-sheet momentum \eqref{pws2}, for example in the zero-winding number case $P_{\rm ws}=0$. To have an arbitrary world-sheet momentum one has to consider string configurations whose 
target space-time image is an open string with end points moving in unison so that $\De x^-$ remains constant. This however leads to the dependence of the target space-time energy $E$ and momentum $J$ on $\a$. Indeed, taking into account that $\De x^+=0$, one finds (assuming $P_{\rm ws}=0$)
\bal
\D t=-({1\ov2}+\a)\D x^-\,,\quad \D \p=({1\ov2}-\a)\D x^-\,,
\eal
and, therefore, different values of $\a$ correspond to different open string configurations. If the gauge-fixed model does not possess any symmetry relating the world-sheet energy and momentum then there seems to be no simple relation between the spectra of the undeformed and deformed models. However, if the models are Lorentz invariant on the world-sheet then the relation is relatively easy to find.

\medskip

We assume again that a model is integrable but the total world-sheet momentum of a state does not vanish. We denote $\cP_\a\equiv P_{\rm ws}$. The spectrum of energies $\cE_\a(R,\cP_\a)$ is determined by Bethe equations \eqref{BE1} where we have $ \sum_{k=1}^N p_k=\cP_\a$. We 
 rewrite \eqref{BE1} in the form which generalises \eqref{BE2}
\bal\la{BE3}
&e^{ip_i R_0 + i\a \cP_\a\w_i}\prod_{k\neq i}^N e^{i\a(p_i\w_k-p_k\w_i)}S_{ik}(\theta_{ik})=e^{im_i(R_0\sinh\theta_i  + \a \cP_\a\cosh\t_i)}\prod_{k\neq i}^N S_{ik}^{(0)}(\theta_{ik})=1\,,\\
&\cE_\a(R,\cP_\a)= \sum_{k=1}^N m_k\cosh\theta_k \,,\quad   \cP_\a=\sum_{k=1}^N m_k\sinh\theta_k\,.
\eal
where we used Lorentz invariance, and that $R=R_0+\a\, \cE_\a(R,\cP_\a)$. Let us now introduce the shifted rapidity $\vt=\t-\Delta\t$ where $\D\t$ is found from the equation
\bal\la{R0Pwsal}
\cR_0\sinh(\theta-\Delta\t)=R_0\sinh\theta + \a \cP_{\a} \cosh\t\quad \Rightarrow\\
R_0=\cR_0\cosh\Delta\t\,,\quad \a \cP_{\a} = -\cR_0\sinh\D\t\,,
\eal
\bal\la{R0al}
R_0^2-\a^2 \cP_{\a}^2 =\cR_0^2\,.
\eal
Eq.\eqref{R0al} shows how the target space-time charge $(J+E)/2=R_0$ depends on the world-sheet momentum and the gauge parameter $\a$. In terms of $\vt_i$  
\eqref{BE3} takes the standard form of the Bethe equations of the undeformed theory on a circle of circumference $\cR_0$
\bal\la{BE4}
&e^{i\cR_0m_i\sinh\vartheta_i}\prod_{k\neq i}^N S_{ik}^{(0)}(\vartheta_{ik})=1\,,\\
\cE_0(\cR_0,\cP_0)&= \sum_{k=1}^N m_k\cosh\vartheta_k\,,\quad   \cP_{0}=\sum_{k=1}^N m_k\sinh\vartheta_k\,.
\eal
We also find the following relations 
\bal
\cE_\a(R,\cP_\a)=\cE_0(\cR_0,\cP_0)\cosh\D\t + \cP_0\sinh\D\t\,,
\eal 
\bal
\cP_{\a}= \cP_0\cosh\D\t + \cE_0(\cR_0,\cP_0)\sinh\D\t\,,
\eal 
and therefore
\bal
\la{Bureq4}
\cE_\a^2(R,\cP_\a) - \cP_{\a}^2 =\cE_0^2(\cR_0,\cP_0) - \cP_0^2\,.
\eal
Eqs.(\ref{R0Pwsal}-\ref{Bureq4}) represent the integrated form of the inhomogeneous inviscid Burgers equation, see e.g. \cite{Tateo16,Tateo18a}
\bal\la{Bureqinhom}
\pa_\a \cE_\a(R,\cP_\a)+{1\ov 2}\pa_R( \cE^2_\a(R,\cP_\a)-\cP_{\a}^2 )=0\,.
\eal
Note that due to the momentum quantisation $\cP_{\a}=2\pi k/R$, $\cP_{0}=2\pi k/\cR_0$ where $k$ is an $\a$-independent integer. The discussion above strictly speaking applies only for large $R$. For finite $R$ one can use the TBA equations which however have the same dependence on $R$ and $\cP_\a$, and therefore lead to the same Burgers equations, cf. \cite{Tateo16,Tateo18a}.

\medskip

Another way (which we cannot fully justify) to derive the inhomogeneous inviscid Burgers equation is as follows.
We consider a state with momentum $\cP_{\a}$ and energy $\cE_\a(R,\cP_\a)$ of a gauge-fixed model on a cylinder of circumference $R$. Then, we go to the reference frame where the world-sheet momentum is zero. Due to the Lorentz invariance we get
\bal
\la{Bureq2}
\cP_{\a} = \cE_\a(R',0)\sinh\psi\,,\quad \cE_\a(R,\cP_\a)= \cE_\a(R',0)\cosh\psi\,, 
\eal
\bal
\la{Bureq3}
\cE_\a^2(R,\cP_\a) - \cP_{\a}^2= \cE_\a^2(R',0)\,,
\eal
where $R'$ is the circumference of the world-sheet cylinder in the new reference frame which  due to the Lorentz contraction satisfies the equation
\bal\nn
{dR'\ov dR}={1\ov\cosh\psi}\,.
\eal
Now, taking into account that $\cE_\a(R',0)$ satisfies the homogeneous inviscid Burgers equation \eqref{Bureqhom}, one derives \eqref{Bureqinhom}.

\subsection{Scalar fields with arbitrary potential} \la{bosTTb}

As the first example of usefulness of the light-cone gauge approach to the \TTb deformation let us consider the deformation of a sigma-model  of $n$ scalar fields with the action
\bal\la{Sbos} 
S_0 = \int_{-r}^{ r}\, {\rm
d}\s{\rm d}\tau\, \big(-{1\over 2}\eta^{\a\b}\partial_\a x^\mu\partial_\b x^\nu\,G_{\mu\nu}-V(x)\big)\,, 
\eal
where $\eta^{\a\b}=\diag(-1,1)$, and $V$ is an arbitrary potential. In the Hamiltonian formalism this action takes the form
\bal\la{Sbos2} 
S_0 = \int_{-r}^{ r}\, {\rm
d}\s{\rm d}\tau\, (p_\mu\dx^\mu- \cH_0)\,, \quad \cH_0\equiv \cH_x+V(x)\,,
\eal
where $\cH_x$ is given by \eqref{Hx}. We want $S_0$ to be the light-cone gauge-fixed action for $\a=0$ of a string sigma model on $\cM$. To have the two-dimensional Lorentz invariance we need to set the winding number $m$ to 0, and therefore $x^+=\tau$. Then, the only way not to get a square-root type gauge-fixed Hamiltonian is to require that the constraint $C_2$ \eqref{c2} is linear in $p_+$ at $\a=0$, and therefore
\bal\la{Gtphi}
G^{++}\big|_{\a=0}=0=G_{--}\big|_{\a=0}\quad \Longrightarrow\quad G_{t\p}={G_{\p\p}+G_{tt}\ov2}\,,
\eal
where we have used \eqref{Gmm}. Solving the constraint $C_2$ \eqref{c2} for $p_+$ with $G_{t\p}$ given by \eqref{Gtphi}, we find the gauge-fixed Hamiltonian at $\a=0$
\bal
\cH_{\rm ws}\big|_{\a=0}={G_{\p\p}-G_{tt}\ov2}\cH_x - 2{G_{\p\p}+G_{tt}\ov G_{\p\p}-G_{tt}}\,.
\eal
We see that to reproduce $\cH_0$ we have to require
\bal
G_{\p\p}-G_{tt}=2\,,\quad G_{\p\p}+G_{tt}=-V(x)\,.
\eal
Thus, to obtain the \TTb deformation of the model \eqref{Sbos}, we need to use the following target space-time metric
\bal
G_{tt}=-(1+{V\ov2})\,,\quad G_{\p\p}=1-{V\ov2}\,,\quad G_{t\p}=-{V\ov2}\,.
\eal
It is easy to check that under these conditions 
\bal
\det G_{\rm lc} = G_{tt}G_{\p\p}-G_{t\p}^2=-1\,,
\eal
and
\bal\la{Gppli}
G^{--}=2V=-G_{++}\,,\quad G^{++}=2\a(1+\a V)=-G_{--}\,,\quad G^{-+}=1+2\a V=G_{-+}\,.
\eal 
Then, for an arbitrary $\a$ the constraint $C_2$ takes the form
\bal
{C_2\ov2}=\alpha (1+\alpha  V)p_+^2+ (1+ 2\alpha  V)p_+ +\cH_x+V-\alpha(1+\alpha  V) (x'^-)^2\,,
\eal
where $x'^-=- p_\mu x'^\mu$. Solving the constraint for $p_+$, we find the gauge-fixed Hamiltonian
\bal
\cH_{\rm ws}={1\ov\a}-{1\ov 2\tilde\a}-{1\ov2\tilde\a}\sqrt{1-4\tilde\a \cH_x +4\tilde\a^2(x'^-)^2}\,,
\eal
where 
\bal
\tilde\a=\alpha(1+\alpha  V)\,.
\eal
Expanding the Hamiltonian, one gets the expected result for a \TTb deformed model
\bal
\cH_{\rm ws}=\cH_x +V+(\cH_x^2 -V^2- (x'^-)^2)\a +\cO(\a^2)\,.
\eal
To see that the model is Lorentz invariant we find the \TTb deformed Lagrangian
\bal
\cL_{\rm ws}=-{1\ov\a}+{1\ov 2\tilde\a}+{1\ov2\tilde\a}\sqrt{1+2\tilde\a(\dx^2-x'^2) -4\tilde\a^2(\dx^2 x'^2-(\dx x')^2)}\,,
\eal
where
\bal
\dx^2\equiv  G_{\mu\nu}\dx^\mu\dx^\nu\,,\quad x'^2\equiv  G_{\mu\nu}x'^\mu x'^\nu\,,\quad \dx x'\equiv  G_{\mu\nu}\dx^\mu x'^\nu\,.
\eal
Obviously it is Lorentz invariant. 
Expanding the Lagrangian, one gets 
\bal\la{LbosTTb}
\cL_{\rm ws}={1\ov2}(\dx^2-x'^2)-V+\det(T^\b{}_\g)\,\a +\cO(\a^2)\,,
\eal
where 
\bal
T^{\b}{}_\g={\pa\cL_{\rm ws}\ov\pa\partial_\b x^\mu}\partial_\g x^\mu -\delta^{\b}_\g\,\cL_{\rm ws}\,,\quad \det(T^\b{}_\g) = T^{0}{}_{0}T^{1}{}_{1}-T^{0}{}_{1}T^{1}{}_{0}\,,
\eal
\bal
T_{\a\b}=\eta_{\a\r}T^{\r}{}_\b=G_{\mu\nu}\partial_\a x^\mu \partial_\b x^\nu -{1\ov 2}\eta_{\a\b}(\eta^{\g\de}\partial_\g x^\mu\partial_\de x^\nu\,G_{\mu\nu}+2V)\,,
\eal
is the canonical stress-energy tensor of the undeformed model \eqref{Sbos} which for the scalar model coincides with the covariant one. This is the \TTb deformation with \TTb$=\det T^{\b}{}_{\g}=-\det T_{\b\g}$. The \TTb deformed Lagrangian \eqref{LbosTTb} seems to agree with the one guessed in \cite{Tateo18a} (for Euclidean world-sheet theory).

\section{Models with fermions and chiral bosons}\la{TTbfermions}

\subsection{Light-cone gauge for Green-Schwarz type models}\la{lcgf}

To consider \TTb deformed models with fermions and chiral bosons we can use string sigma models of the Green-Schwarz type. 
Let us assume that in addition to bosonic fields $X^M$ we have $n_f$ real fermionic variables $\theta^f$ and $n_c$ real chiral bosons\footnote{We call these bosons chiral because, as we will see, if in an undeformed theory they are free then they satisfy  the equations of motion $\pa_\pm\vp^c=0$. } $\vp^c$ which are world-sheet scalars. We combine the fermions and chiral bosons into 
\bal\la{Psia}
\Psi^a = (\t^f,\vp^c)\,,\quad  \Psi^a=\t^a\,,\ a=1,\ldots,n_f\,,\quad \Psi^a=\vp^a\,,\ a=n_f+1,\ldots,n_f+n_c\,,
\eal
and we call the index $a$ fermionic if $\Psi^a$ is a fermion, and bosonic if it is a boson.

\medskip

We consider  a reasonably general  sigma model with the following action 
\bal \la{Sf1} 
S = -\int_{-r}^{ r}\, {\rm
d}\s{\rm d}\tau\big( {1\over 2}\g^{\a\b}\Pi_\a^M\Pi_\b^N\rG_{MN} + i\e^{\a\b}  \partial_\a X^M\Psi^a \rA_{Mab}\pa_\b \Psi^b + {i\ov 2}\e^{\a\b}  \partial_\a \Psi^a\rF_{ab}\pa_\b\Psi^b  \big)\,,
\eal
where the skew-symmetric Levi-Civita symbol is defined by $\epsilon^{\tau\sigma}=1$, the world-sheet currents $\Pi_\a^M$ are given by
\bal
\Pi_\a^M = \partial_\a X^M + i\,\Psi^a\G^M_{ab} \pa_\a\Psi^b\,,
\eal
and $\rG_{MN}$, $\G^M_{ab}$, $ \rA_{Mab}$ and $\rF_{ab}$ are arbitrary  functions of $x^\mu$ and $\Psi^a$ whose parity depends on whether $a,b$ are fermionic or bosonic, e.g.
\bal
\rG_{MN}=\rG_{MN}(x,\Psi)=G_{MN}(x,\vp) + G_{MN,ab}(x,\vp)\t^a\t^b + G_{MN,abcd}(x,\vp)\t^a\t^b\t^c\t^d +\cdots\,,
\eal
\bal
\G^M_{ab}=\G^M_{ab}(x,\Psi)=\vG^M_{ab}(x,\vp) + \vG^M_{ab,cd}(x,\vp)\t^c\t^d + \vG^M_{ab,cdef}(x,\vp)\t^c\t^d\t^e\t^f +\cdots\,,
\eal
if $a,b$ are both either fermionic or bosonic, and 
\bal
\G^M_{ab}=\G^M_{ab}(x,\Psi)=\vG^M_{ab,c}(x,\vp)\t^c + \vG^M_{ab,cde}(x,\vp)\t^c\t^d\t^e + \cdots\,,
\eal
if the indices are of opposite parity, and similar expansions for $ \rA_{Mab}$ and $\rF_{ab}$. These functions are real\footnote{It is worthwhile noting that the fermions and bosons do not have to be real. All one has to assume is that $\Psi^\dagger$ is related to $\Psi$ by means of a charge conjugation matrix $C$. Then, the reality of the action \eqref{Sf1} would lead to reality conditions on these functions which would depend on $C$. The considerations below would not be modified and the final action would still have the same form  \eqref{Sf4}.}
$(\rG_{MN})^*=\rG_{MN}$, $(\G^M_{ab})^*=\G^M_{ab}$ if at least one of the indices $a,b$ is fermionic, and imaginary if both indices are bosonic. This follows from the conjugation rule for  fermions: $(\theta^a\theta^b)^*=(\theta^b)^*(\theta^a)^* = -\theta^a\theta^b$. Note that  $\rF_{ab}$ is  symmetric under the exchange of $a,b$ if $a,b$ are fermionic, and it is skew-symmetric if at least one of the indices $a,b$ is bosonic.

\medskip

Obviously, as in the purely bosonic case, the action is invariant under the shifts of $t$ and $\p$. We do not assume any target space symmetry, and therefore $\G^M_{ab}$  are not in general related to gamma matrices. Then, we assume that the reparametrisation invariance is the only gauge symmetry of the model, and therefore it is not $\k$-invariant.\footnote{If one starts with a $\k$-invariant model, then one first imposes a kappa-symmetry gauge condition and reduces its action to the form \eqref{Sf1}.}

\medskip

Introducing auxiliary  momentum-like variables $\bp_M$,  we rewrite the
 action (\ref{Sf1}) in the first-order form 
\bal \la{Sf2} 
S=\int_{- r}^{ r}\, {\rm d}\s{\rm d}\tau\, \Big(& \bp_M \Pi_0^M +
{\g^{01}\ov\g^{00}} C_1+ {1\ov 2\, \g^{00}}C_2\\
&- i\e^{\a\b}  \partial_\a X^M\Psi^a \rA_{Mab}\pa_\b \Psi^b  - {i\ov 2}\e^{\a\b}  \partial_\a \Psi^a\rF_{ab}\pa_\b\Psi^b\Big)\,.  
\eal
Here $\bp_M$ satisfy the equations 
 \bea\nonumber 
 \bp_M ={\de S\ov \de \Pi_0^M} = -\g^{0\b}\Pi_\b^N\,\rG_{MN}\,,
\eea and
$C_1$ and $C_2$ are the two Virasoro constraints 
\bea \nonumber 
C_1=\bp_M\Pi_1^M\,,\quad
C_2=\rG^{MN} \bp_M \bp_N + \Pi_1^M\Pi_1^N \rG_{MN}\,.
\eea
Collecting the terms with $\dot X^M$ and $\dot\Psi^a$, we bring the action \eqref{Sf2} to the form
\bal \la{Sf3} 
S=\int_{- r}^{ r}\, {\rm d}\s{\rm d}\tau\, \left( p_M \dot X^M + i\,p^\Psi_b\dot\Psi^b +
{\g^{01}\ov\g^{00}} C_1+ {1\ov 2\, \g^{00}}C_2\right)\,,
\eal 
where $p_M$ is the momentum  conjugated to $X^M$ and related to $\bp_M$ as 
\bal\la{pmpim}
p_M= \bp_M- i\,\Psi^a \rA_{Mab}\Psi'^b \,,
\eal
and $p^\Psi_b$  is  given by
\bal
p^\Psi_b=\Psi^a\bp_M\G^M_{ab} +X'^M\Psi^a  \rA_{Mab} +\Psi'^a \rF_{ab}\,.
\eal
Now, we can proceed in the same way as in section \ref{blc}. 
We introduce the light-cone
coordinates and momenta \eqref{lcc}, and rewrite the
action (\ref{Sf3}) in the form \eqref{S3} 
with 
\bal 
\la{Cf1b} C_1=\bp_+\Pi_1^+
\,+\, \bp_-\Pi_1^- \,+\, \bp_\mu \Pi_1^\mu\,  ,
\eal 
\bal \la{fc2}
C_2 &= \rG^{--} \bp_-\bp_-\,  +\, 2\rG^{-+}\bp_-\bp_+ \, +\,
\rG^{++}\bp_+\bp_+\\ 
&+
\rG_{--} \Pi_1^-\Pi_1^-\,  +\,
2\rG_{-+} \Pi_1^-\Pi_1^+
+ \rG_{++}\Pi_1^+\Pi_1^+ \, + \,
2\cH_x\,,
\eal 
and
\bal\la{fHx} 
\cH_x = {1\ov2}\big(\rG^{\mu\nu}\bp_\mu \bp_\nu +  \Pi_1^\mu \Pi_1^\nu\,
\rG_{\mu\nu}\big)\,.
\eal 
Here $\bp_M$ are related to $p_M$ by  \eqref{pmpim}, and the light-cone components of the target space metric are given by
(\ref{Gpp}-\ref{Gmp}) with $G\to\rG$.

\medskip

Next, we impose the uniform light-cone gauge \eqref{ulc} with the winding number $m=0$  for simplicity. Then, we 
solve the Virasoro
constraint  $C_1$ for $x'^-$ 
\bal\la{Cf1c}
 C_1\,&=\,0\, \quad \Rightarrow \quad
 \Pi_1^-\,=\,  - \, {\bp_+\Pi_1^++
\bp_\mu \Pi_1^\mu\ov \bp_-}\quad \Rightarrow \\
 x'^-\,&=\, -i\Psi^a \G^-_{ab}\Psi'^b  - \, {i\bp_+\Psi^a \G^+_{ab}\Psi'^b +
\bp_\mu \Pi_1^\mu\ov \bp_-}\,,
\eal
where
\bal\la{bpM0}
 \bp_-&=1+ i\,\Psi^a\rA_{-ab} \Psi'^b \,,\quad \bp_+=p_++ i\,\Psi^a \rA_{+ab}\Psi'^b \,,\quad \bp_\mu=p_\mu+ i\,\Psi^a \rA_{\mu ab}\Psi'^b \,,\\
 \Pi_1^-&=x'^- +i\Psi^a \G^-_{ab} \Psi'^b \,,\quad \Pi_1^+= i\,\Psi^a \G^+_{ab}\Psi'^b \,.
\eal
Note that $x'^-$ is a polynomial in $\t$'s and $\t'$'s. Finally, we substitute the solution
into $C_2$,  and solve the resulting equation   for $\bp_+$. The string action
(\ref{Sf3})  then takes the gauge-fixed form
\bea \la{Sf4} 
S_\a=\int_{- r}^{ r}\, {\rm d}\s{\rm d}\tau\, \left( p_\mu \dot{x}^\mu+ i\,p^\Psi_b\dot\Psi^b
\,-\, \cH_{\rm ws} \right)\, ,\quad
  \cH_{\rm ws} \,=\, -p_+(p_\mu, x^\mu ,x'^\mu,\Psi^a,\Psi'^a )\,,
 \eea 
where 
\bal\la{bpM}
-p_+=-\bp_++ i\,\Psi^a \rA_{+ab} \Psi'^b\,,
\eal
\bal\la{pthb}
p^\Psi_b=\bp_+\Psi^a\G^+_{ab}+\bp_-\Psi^a\G^-_{ab} +x'^-\Psi^a  \rA_{-ab} +\bp_\mu\Psi^a\G^\mu_{ab} +x'^\mu\Psi^a  \rA_{\mu ab} +\Psi'^a \rF_{ab}\,.
\eal
This is the \TTb deformed action of the model with the action $S_0$. It has a nontrivial Poisson structure. To see how one can handle such a structure efficiently see \cite{AAF, FPZ06,AFrev}.

\medskip

Let us also mention that one can choose $\G^\p_{ab}$, $\rA_{-ab} $, and so on, in such a way that some of the fermions and bosons in $\Psi$ would be non-dynamical. Then, one may want to eliminate the non-dynamical fields to deal only with the dynamical ones. In fact, it seems that with the help of non-dynamical fields one can set all transversal fields $x^\mu$ to zero, and still get the \TTb deformation of any model. In particular, the bosonic model discussed in the previous section can be obtained in such a way.

\subsection{Scalars, fermions and chiral bosons with arbitrary potential}

The formalism developed in the previous subsection can be used to derive a \TTb deformed action of a sigma model  of $n_b$ scalar fields $x^\mu$,  and $n_f^-$ right-moving real fermions $\t_-^i$, and $n_f^+$ left-moving real fermions $\t_+^r$, and $n_c^-$ right-moving real chiral bosons $\vp_-^i$, and $n_c^+$ left-moving real chiral bosons $\vp_+^r$. For Lorentz invariant models the fermions are world-sheet spinors, while the numbers of chiral bosons should be even and half of them should be world-sheet scalars and the other half world-sheet chiral vectors. We combine fermions and chiral bosons again into two fields
\bal\la{Psib}
\Psi^a_\pm = (\t^f_\pm,\vp^c_\pm)\,,\quad  \Psi^a_\pm=\t^a_\pm\,,\ a=1,\ldots,n_f^\pm\,,\quad \Psi^a_\pm=\vp^a_\pm\,,\ a=n_f^\pm+1,\ldots,n_f^\pm+n_c^\pm\,,
\eal
and consider the following undeformed action
\bal\la{Sbosf} 
S_0 = \int_{-r}^{ r}\, {\rm
d}\s{\rm d}\tau\, (-{1\over 2}\eta^{\a\b}\partial_\a x^\mu\partial_\b x^\nu\,\rG_{\mu\nu}+i\Psi_-^i\rK_{ij}^+\pa_+\Psi_-^j+i\Psi_+^r\rK_{rs}^-\pa_-\Psi_+^s-\rV)\,.
\eal
Here 
\bal
\pa_\pm=\pa_\tau\pm\pa_\s\,,\ i,j=1,\ldots,n_f^-+n_c^-\,, \ r,s=n_f^-+n_c^-+1,\ldots,n_f^-+n_c^-+n_f^++n_c^+\,,
\eal
$\rK_{ab}^\pm$  are real if at least one of the indices $a,b$ is fermionic, and imaginary if both indices are bosonic. Then, $\rV$ is an arbitrary potential. They all depend on $x$ and $\Psi$ (in a Lorentz-invariant way if necessary). In the first-order formalism this action takes the form
\bal\la{Sbosf2} 
&S_0 = \int_{-r}^{ r}\, {\rm
d}\s{\rm d}\tau\, (p_\mu\dx^\mu+i\Psi_-^i\rK_{ij}^+\dot\Psi_-^j+i\Psi_+^r\rK_{rs}^-\dot\Psi_+^s- \cH_0)\,, \\
& \cH_0\equiv \cH_x+\rV -i\Psi_-^i\rK_{ij}^+\Psi_-'^j+i\Psi_+^r\rK_{rs}^-\Psi_+'^s\,,
\eal
where $\cH_x$ is given by \eqref{Hx}. We want $S_0$ to be the light-cone gauge-fixed action for $\a=0$ of a string sigma model on $\cM$ with $n_f=n_f^-+n_f^+$ fermions $\t^a=(\t^i_-,\t^r_+)$, and $n_c=n_c^-+n_c^+$ chiral bosons $\vp^a=(\vp^i_-,\vp^r_+)$, so that 
$\Psi^a=(\Psi^i_-,\Psi^r_+)$. It is easy to see that just as for the bosonic case discussed in section \ref{bosTTb}  to reproduce $\cH_x +\rV$ we have to use the same target space-time metric
\bal
\rG_{tt}=-(1+{\rV\ov2})\,,\quad \rG_{\p\p}=1-{\rV\ov2}\,,\quad \rG_{t\p}=-{\rV\ov2}\,.
\eal
\bal\la{Gpplif}
\rG^{--}=2\rV=-\rG_{++}\,,\quad \rG^{++}=2\a(1+\a \rV)=-\rG_{--}\,,\quad \rG^{-+}=1+2\a \rV=\rG_{-+}\,.
\eal 
Then, the gauge-fixed Hamiltonian becomes
\bal\la{Hfws0}
\cH_{\rm ws}\big|_{\a=0}=\bp_-\rV+{\bp_-\cH_x\ov \bp_-^2-(\Pi_1^+)^2} -{\bp_\mu\Pi_1^\mu\Pi_1^+ \ov \bp_-^2-(\Pi_1^+)^2} + i\,\Psi^a \rA_{+ab}\Psi'^b \,.
\eal
Comparing \eqref{Hfws0} with $\cH_0$, we see that we have to impose the conditions
\bal\la{Atab}
\bp_-\big|_{\a=0}&=1\quad \Rightarrow\quad  \rA_{-ab}\big|_{\a=0}= {1\ov2}(\rA_{\p ab}- \rA_{tab})=0\quad \Rightarrow\\
\rA_{+ab}&= 2\rA_{\p ab}\,,\quad \rA_{-ab}=-\a\rA_{+ab}= -2\a\rA_{\p ab}\,,
\eal
\bal\la{Gtab}
\Pi_1^+\big|_{\a=0}&=0\quad \Rightarrow\quad  \G^+_{ab}\big|_{\a=0}={1\ov2}(\G^\p_{ab}+\G^t_{ab})=0\quad \Rightarrow\\
\G^-_{ab}&=2\G^\p_{ab} \,,\quad \G^+_{ab}=\a\G^-_{ab}=2\a\G^\p_{ab}\,,
\eal
\bal\la{Atab2}
\rA_{+ij}=2\rA_{\p ij}=-\rK_{ij}^+\,,\quad \rA_{+rs}=2\rA_{\p rs}=\rK_{rs}^-\,,\quad \rA_{\p ir}=\rA_{\p ri}=0\,.
\eal
Next we need to equate $p_b^\Psi\dot\Psi^b\big|_{\a=0}$ with the kinetic term in \eqref{Sbosf2}. We get
\bal\nn
p_b^\Psi\dot\Psi^b\big|_{\a=0}=(\Psi^a\G^-_{ab}  +\bp_\mu\Psi^a\G^\mu_{ab} +x'^\mu\Psi^a  \rA_{\mu ab} +\Psi'^a \rF_{ab})\dot\Psi^b =\Psi_+^i\rK_{ij}^+\dot\Psi_+^j+\Psi_-^r\rK_{rs}^-\dot\Psi_-^s\,,
\eal
which leads to the solution
\bal\la{Gpabf}
\G^-_{ij}=2\G^\p_{ij}= \rK_{ij}^+\,,\quad \G^-_{rs}=2\G^\p_{rs}=\rK_{rs}^-\,,\quad \G^\p_{ir}=\G^\p_{ri}=0\,,
\eal
\bal\la{Gmuab}
\G^\mu_{ab}= \rA_{\mu ab}= \rF_{ab}=0\,.
\eal
Calculating 
 the constraint $C_2$ for an arbitrary $\a$ with $\Pi_1^-$ given by \eqref{Cf1c}, one gets
\bal
{C_2\ov2}&=\alpha (1+\alpha  \rV)(1-{(\Pi_1^+)^2\ov \bp_-^2})\bp_+^2\\
&+ \bp_-\Big(1+ 2\alpha\rV-2 \a(1 + \a\rV) {\bp_\mu\Pi_1^\mu \Pi_1^+\ov \bp_-^3} 
- (1 + 2 \a \rV){(\Pi_1^+)^2\ov \bp_-^2}\Big)\bp_+\\
&+\cH_x+\bp_-^2\rV-\alpha(1+\alpha  \rV) {(\bp_\mu\Pi_1^\mu)^2\ov \bp_-^2}-(1+2\alpha  \rV) {\bp_\mu\Pi_1^\mu \Pi_1^+\ov \bp_-}  - \rV (\Pi_1^+)^2\,.
\eal
Solving the constraint for $\bp_+$, one gets
\bal
-\bp_+={\bp_-\ov \a}
-\frac{\bp_-}{2 \tilde\a}-\frac{p_\mu x'^\mu
  \Pi_1^+}{\bp_-^2-(\Pi_1^+)^2}
-\frac{\bp_-}{2 \tilde\a} \sqrt{1-4\tilde\b\cH_x+4 \tilde\b^2(p_\mu x'^\mu)^2}\,,
\eal
where 
\bal
\tilde\a=\alpha(1+\alpha  \rV)\,,\quad \tilde\b=\frac{\tilde\a
 }{\bp_-^2-(\Pi_1^+)^2}\,,
\eal
 and  due to (\ref{Atab}-\ref{Gmuab})
\bal\la{bpM3}
 \bp_-&=1-i\,\a\Psi^a \rA_{+ ab}\Psi'^b \,,\quad  
 \Pi_1^+= i\,\a\Psi^a \G^-_{ab}\Psi'^b \,,
\quad
\bp_\mu=p_\mu\,,\quad \Pi_1^\mu=x'^\mu\,. 
\eal
The gauge-fixed Hamiltonian is then given by
\bal
\cH_{\rm ws}=-\bp_++ i\,\Psi^a  \rA_{+ ab}\Psi'^b\,,
\eal
while the gauge-fixed Lagrangian in the first-order form is
\bal\la{Lwsf1}
\cL_{\rm ws}=p_\mu \dot{x}^\mu+ i\,p^\Psi_b\dot\Psi^b
\,-\, \cH_{\rm ws} \,,
\eal
where
\bal\la{pthb2}
p^\Psi_b=\bp_-\Psi^a\G^-_{ab} +\a\Psi^a\G^-_{ab}\bp_+-\a x'^-\Psi^a  \rA_{+ab} \,,
\eal
\bal
x'^-\,&=\, - \, {
p_\mu x'^\mu\ov \bp_-}-i\Psi^a\G^-_{ab}\Psi'^b - \, i\a{\Psi^a \G^-_{ab}\Psi'^b \ov \bp_-}\bp_+\,.
\eal
The equations of motion for $p_\mu$ take the form
\bal
\dx^\mu+Q x'^\mu=Y{p^\mu-2\tilde\b p_\nu x'^\nu x'^\mu\ov \sqrt{1-4\tilde\b\cH_x+4 \tilde\b^2(p_\mu x'^\mu)^2}}\,,
\eal
where
\bal
Y={\tilde\b\ov\tilde\a}\bp_-(1+i\a \Psi^a \G^-_{ab}\dot\Psi^b +i\a {\Pi_1^+\ov\bp_-}\Psi^a \rA_{+ ab}\dot\Psi^b )\,, 
\eal
\bal
Q={i\a\ov\bp_-}\Psi^a \rA_{+ ab}\dot\Psi^b +{\Pi_1^+\ov\bp_-}Y\,, 
\eal
which can be used to eliminate $p_\mu$ from the Lagrangian \eqref{Lwsf1}. 

\medskip

To write the resulting Lagrangian in a compact form  let us introduce the following notations
\bal
\rK^+_\g\equiv \Psi_-^i\rK_{ij}^+\pa_\g\Psi_-^j\,,\quad \rK^-_\g\equiv \Psi_+^r\rK_{rs}^-\pa_\g\Psi_+^s\,,
\eal
in terms of which we have
\bal
\Psi^a \G^-_{ab}\pa_\g\Psi^b  = \rK^+_\g+\rK^-_\g\,,\quad \Psi^a \rA_{+ab}\pa_\g\Psi^b = -\rK^+_\g+\rK^-_\g\,.
\eal
We also use light-cone coordinates on the world-sheet
\bal
\pa_\pm x^\mu =\pa_\tau x^\mu\pm \pa_\s x^\mu\,.
\eal
Then, the gauge-fixed Lagrangian takes the form
\bal\la{LfermTTb}
\cL_{\rm ws}&={\, i\rK^+_++i\rK^-_-+\a(\rK^+_-\rK^-_+-\rK^+_+\rK^-_-)\, \ov 2(1+\a\rV) }\,-\,{1\ov\a}+{1\ov2\tilde\a}+{1\ov2\tilde\a}\sqrt{\Lambda}\,,
\eal
where 
\bal
\Lambda&=\big(1 + i \a (\rK^-_- + \rK^+_+) + \a^2 (\rK^+_- \rK^-_+ - \rK^-_- \rK^+_+)\big)^2
\\
& +2 \tilde\a\pa_-x\pa_+x \big(1 + i \a (\rK^-_- + \rK^+_+) -\a^2 (\rK^+_- \rK^-_+ + \rK^-_- \rK^+_+)\big)  
\\
&+ \tilde\a^2\big((\pa_-x\pa_+x)^2-(\pa_-x)^2 (\pa_+x)^2\big) 
- 2 i \a\tilde\a \big((\pa_-x)^2 \rK^-_+ + \rK^+_- (\pa_+x)^2\big) 
\\
& + 2 \a^2\tilde\a \big((\pa_-x)^2 \rK^-_+ \rK^+_+ + \rK^-_- \rK^+_- (\pa_+x)^2\big) \,,
\eal
and
\bal
\pa_-x\pa_+x\equiv G_{\mu\nu}\pa_-x^\mu\pa_+x^\nu\,,\quad (\pa_\pm x)^2\equiv G_{\mu\nu}\pa_\pm x^\mu\pa_\pm x^\nu\,.
\eal
In the derivation of  \eqref{LfermTTb} we have never needed Lorentz invariance of the undeformed model. Let us assume that $\rV$, $\rK^+_+$ and $\rK_-^-$ are Lorentz scalars while $\rK^+_-$ transforms as $(\pa_-x)^2$ and $\rK^-_+$ as $(\pa_+x)^2$. Then, one sees from \eqref{LfermTTb} that this Lagrangian is  Lorentz invariant. It is also clear that if $\rK_{ab}^\pm$ are Lorentz scalars, then $\t_-^i$ and $\t_+^r$ are the world-sheet right and left Majorana-Weyl spinors, respectively. It is certainly well-known that in the Green-Schwarz sigma models, as a result of the light-cone gauge fixing, the world-sheet fermion scalars become the world-sheet spinors.  To understand what happens with the chiral bosons let us consider the field-independent part of $\rK_{ab}^\pm$ which is necessarily skew-symmetric\footnote{The symmetric part of $\rK_{ab}^\pm$ always describes an interaction of the chiral bosons between themselves and other fields.} for bosonic indices. Then, by an orthogonal transformation it can be brought to the standard symplectic form
\bal
iK_{ab}^\pm = 
\left(
\begin{array}{ccc}
  +1& {\rm if}  &  a=b+{n_c^\mp/2} \\
  -1&  {\rm if}  &  b=a+{n_c^\mp/2}  \\
  0&  {\rm otherwise} &   
\end{array}
\right)\,.
\eal
Thus, up to quadratic order in the chiral bosons the bosonic parts of $K^+_+$ and $K_-^-$ take the form
\bal
iK^+_+ &= \sum_{i=1}^{n_c^-/2}(A_-^i\pa_+\vp_-^i-\vp_-^i\pa_+A_-^i)\,,\quad A_-^i=\vp_-^{i+n_c^-/2}\,,\\
iK^-_- &= \sum_{i=1}^{n_c^+/2}(A_+^i\pa_-\vp_+^i-\vp_+^i\pa_-A_+^i)\,,\quad A_+^i=\vp_+^{i+n_c^+/2}\,,
\eal
which shows that we can take $\vp_\pm^i$ and $A_\pm^i$, $i=1,\ldots, n_c^\pm/2$ to be world-sheet  scalars and vectors, respectively.

\medskip
 
Expanding the Lagrangian \eqref{LfermTTb}, one gets 
\bal\la{LfermTTb2}
\cL_{\rm ws}={1\ov2}\pa_-x\pa_+x+ i\rK^+_++i\rK^-_--\rV +\det(T^\b{}_\g)\,\a +\cO(\a^2)\,,
\eal
where 
\bal
T^{\b}{}_\g={\pa\cL_{\rm ws}\ov\pa\partial_\b x^\mu}\partial_\g x^\mu+{\pa\cL_{\rm ws}\ov\pa\partial_\b \Psi^a_+}\partial_\g \Psi^a_++{\pa\cL_{\rm ws}\ov\pa\partial_\b \Psi^a_-}\partial_\g \Psi^a_- -\delta^{\b}_\g\,\cL_{\rm ws}\,,\
\eal
is the canonical stress-energy tensor of the undeformed model \eqref{Sbos} which for the fermion model does not coincide with the covariant one. This agrees with the consideration of \TTb deformed supersymmetric systems of free massless bosons and fermions in \cite{Sfondrini18b,Sfondrini19a}. 
Explicitly, one gets
\bal\la{Tcanf}
T^{-}{}_-&=\rV-i\rK^+_+\,,\quad T^{+}{}_+=\rV-i\rK^-_-\,,\\
 T^{+}{}_-&={1\ov2}(\pa_-x)^2+i\rK^+_-\,,\quad T^{-}{}_+={1\ov2}(\pa_+x)^2+i\rK^-_+\,,
\eal
which is not symmetric off-shell (we lower indices with $\eta_{+-}=\eta_{-+}=1/2$). For Lorentz invariant models the canonical and covariant tensors differ by terms  vanishing on-shell, therefore there is an $\a$-dependent nonlinear field redefinition which transforms $\det T_{\rm can}$ to $\det T_{\rm cov}$.\footnote{I thank Alessandro Sfondrini for a discussion of this point.} This redefinition  can be found as a series in $\a$ but it would ruin the nice structure of the Lagrangian \eqref{LfermTTb}. It is also unclear if this redefinition can be implemented in quantum theory.

\medskip

If there are no bosonic $x^\m$ fields the Lagrangian \eqref{LfermTTb} simplifies drastically, and one gets
 \bal\la{LfermTTb0}
\cL_{\rm AAF}&={\, i\rK^+_++i\rK^-_-+\a(\rK^+_-\rK^-_+-\rK^+_+\rK^-_-)\, - \rV \ov 1+\a\rV }\,.
\eal  
If in addition there are no chiral bosons then the expansion in powers of $\rV$ terminates, and one gets
a generalisation of the AAF model \cite{AAF} to any number of  fermions and any potential. 
If the undeformed model is integrable then the AAF model is integrable too.  A \TTb deformation of the massive Thirring model was discussed in \cite{Bonelli18}. However,  the covariant stress-energy tensor was used there, and by this reason the Lagrangian obtained in \cite{Bonelli18} differs from \eqref{LfermTTb0}, and has a much more complicated structure.

\section{Conclusions}

In this paper we have developed the light-cone gauge approach to the \TTb deformed two-dimensional models which may or may not be Lorentz invariant. To demonstrate the power of this approach we have found the \TTb deformed Lagrangian \eqref{LfermTTb} of a very general system of scalars, fermions and chiral bosons with an arbitrary potential. Most of the \TTb deformed models studied before \cite{Tateo16,Bonelli18,Tateo18a,Sfondrini18b,Sfondrini19a,Sethi18} are  particular cases of this system. Let us mention some open questions and generalisations of the approach.

\medskip

It is obvious that any two gauges of a string sigma model are related to each other by a field-dependent transformation of coordinates which is just a consequence of the reparametrisation invariance of the model. To our knowledge a change of coordinates relating uniform  light-cone gauges with different gauge parameters is not known explicitly. It would be interesting to find it, and compare with the transformation  in \cite{Tateo18b}. We suspect that the canonical Noether stress-energy tensor should be used for models with fermions in their formula.
  
\medskip
 
It was argued in \cite{Dubovsky17,Dubovsky18} that a \TTb deformed model can be obtained if one couples an undeformed Lorentz invariant theory to the flat space Jackiw-Teitelboim (JT)  gravity.
It would be of interest to understand the relation of our approach to the JT one. In particular, the light-cone gauge approach is democratic -- it is a matter of convenience which model is called undeformed and which one its deformation. Moreover, Lorentz  invariance is not important at all. On the other hand, the JT approach distinguishes  an undeformed model and, as far as we can see, requires Lorentz  invariance.
 
\medskip

Obviously, if an undeformed model is integrable the \TTb deformed model is integrable too. The simplest way to find a \TTb deformed Lax pair  is to start with a reparametrisation invariant Lax pair for the underlying string sigma model, impose the light-cone gauge, express unphysical fields and two-dimensional metric in terms of physical fields,  and substitute the expressions into the Lax pair. 
That was how Lax pairs for the bosonic part of the light-cone \ads superstring and the AAF model were found \cite{AF04,AAF}. If a Lax pair of the undeformed model is known then there should exist a way to get from it  a reparametrisation invariant Lax pair. 
  
\medskip

As immediately follows from the light-cone gauge approach if an undeformed model is supersymmetric the \TTb deformed model is supersymmetric too but the supersymmetry may be realised nonlinearly, see \cite{Sfondrini18b,Sfondrini19a,Sethi18}.  
However, in the case of supersymmetric models it might be more natural to use the NSR strings coupled to two-dimensional supergravity. The super-Virasoro constraints would require to impose additional fermionic gauge conditions which would lead to extra gauge parameters. This may allow one to consider more general multi-parameter deformations which involve supercharges. 

\medskip

 In the case of integrable models the \TTb deformation belongs to a class of more general higher-spin deformations introduced in \cite{SZ16}.  It is natural to expect that they can be described in the light-cone gauge approach if one considers strings coupled to W gravity, see \cite{Hull93} for a review. 
 
\medskip

The light-cone gauge approach seems to be the most natural way to study \TTb type deformations which break Lorentz invariance of an undeformed model, the simplest example being the $J\bar{T}$    deformation \cite{Guica17}, and its higher-spin generalisation recently discussed in \cite{Tateo19}.  In this case the undeformed model possesses an additional conserved $U(1)$ current $J$ which from the string sigma model point of view would be a consequence of an additional space isometry in one of the transverse directions. It seems one way to get the $J\bar{T}$    deformation is to include the third isometry direction and its conjugate momentum in the light-cone gauge definition. This would produce a multi-parameter gauge condition, and the corresponding gauge-fixed model would be a multi-parameter deformation by \TTb,  $J\bar{T}$ and $\bar J{T}$ operators. Another way would be to couple the $U(1)$ current to a non-dynamical gauge field. 
 
\medskip

 It is worthwhile mentioning that some non-Lorentz invariant models can be obtained from the action \eqref{Sf1} if one uses non-dynamical fermions and bosons, see the paragraph at the very end of subsection \ref{lcgf}. In particular, it is easy to derive the \TTb deformed action for nonlinear matrix Schr\"odinger model which appears to be of the Nambu-Goto form too.
In this paper we assumed that $G_{t\mu}=G_{\p\mu}=0$ which seems to be a necessary condition for Lorentz invariance.
If $G_{t\mu}$ and $G_{\p\mu}$ do not vanish then Lorentz invariance in general is broken. It might be interesting 
to analyse what kind of non-Lorentz invariant models one can get in this case, in particular if one can reproduce the results of \cite{Cardy18}. 

\medskip

The consideration in this paper was purely classical. Obviously, a \TTb deformed model is not renormalisable, and quantisation of such a model leads to infinitely many quantum versions corresponding to one and the same classical model. The quantum \TTb deformed model by definition is the one whose spectrum satisfies the  inhomogeneous inviscid Burgers equation. It is unclear whether a \TTb quantisation scheme always exists. Even if the model is integrable this may not be the most natural quantisation. For example, the two-particle S-matrix of the AAF model \cite{AAF} was calculated in \cite{KZ06} in a particular quantisation scheme, and it was shown to be factorisable  in \cite{Melikyan11}. The S-matrix, however, is not equal to the \TTb CDD factor, and leads to a nontrivial spectrum. It would be interesting to find the \TTb quantisation scheme for the AAF model, and to understand in general for which models such a scheme exists.

\medskip

Assuming the quantum \TTb deformed model exists, the next question is how to compute its form factors. It is a very hard problem, and it seems the only way to solve it is to understand in full detail the relation between two light-cone gauges with different gauge parameters. In particular it might be necessary to understand how the change of coordinates discussed above is implemented on the quantum level.

\section*{Acknowledgements}

I would like to thank Tristan McLoughlin and Alessandro Sfondrini for fruitful discussions, and 
Alessandro Sfondrini for useful comments on the manuscript. 



\end{document}